\documentclass[epj]{svjour}
\usepackage{graphics}
\usepackage{amssymb}
\begin{document}
\title{Integral and derivative dispersion relations, analysis of the forward scattering data.}
\author{E. Martynov\thanks{\emph{on leave from the} Bogolyubov Institute for
Theoretical Physics, National Academy of Sciences of Ukraine, 03143
Kiev-143, Metrologicheskaja 14b, Ukraine} \and J.R. Cudell \and O.V.
Selyugin\thanks{\emph{on leave from the} Bogoliubov Theoretical
Laboratory, JINR, 141980 Dubna, Moscow Region, Russia}
%
}                     
%
%
\institute{Institut de Physique, Universit\`{e} de Li\`{e}ge,B\^{a}t.
B5-a, Sart Tilman, B4000 Li\`{e}ge, Belgium }
\date{}
%
\abstract{ Integral and derivative dispersion relations (DR) are
considered for the forward scattering $pp$ and $\bar pp$ amplitudes. A new
representation for the derivative DR, valid not only at high energy, is
obtained. The data on the total cross sections for $pp (\bar pp)$
interaction as well as the data on the parameter $\rho $ are analyzed
within the various forms of the DR and high-energy Regge models. It is
shown that three models for the Pomeron, Simple pole Pomeron, Tripole
Pomeron and Dipole Pomeron (the both with the intercept equal unit) lead
to practically equivalent description of the data at $\sqrt{s}>$5 GeV. It
is also shown that the correctly calculated low-energy part of the
dispersion integral (from the two-proton threshold up to $\sqrt{s}=$5 GeV)
allows to reproduce well the $\rho $ data at low energies without
additional free parameters.
\PACS{
      {13.85.Lg}{}   \and
      {11.55.Fv}{}   \and
      {11.55.Jy}{}
     } 
} 
\maketitle
\section{Introduction}
\label{intro} The energy dependence of the hadronic total cross sections
as well as that of the parameters $\rho=\Re eA(s,0)/\Im mA(s,0)$ - the
ratios of the real to the imaginary part of the forward scattering
amplitudes - was widely discussed quite a long time ago (see \cite
{rhohist,DDR} and references therein). However, in spite of recent
detailed investigations on the subject \cite{COMPETE}, the theoretical
situation remains somewhat undecided, mainly because of the $\rho$
parameter.

In the papers \cite{COMPETE}, all available data on $\sigma_{tot}(s)$ and
$\rho(s)$ for hadron-hadron, photon-hadron and photon-photon interactions
were considered. Many analytical models for the forward scattering
amplitudes were fitted and compared. The ratio $\rho$ was calculated in
explicit form, from the imaginary part parametrised by contributions from
the pomeron and secondary reggeons. The values of the free parameters were
determined from the fit to the data at $s\ge s_{min}$, where
$\sqrt{s_{min}}=5$ GeV. Omitting all details, we note here the main two
conclusions. The best description of the data is obtained for the model
with $\sigma_{tot}$ rising as $\log^{2}s$. The model with
$\sigma_{tot}(s)\propto s^{\epsilon}, \epsilon >0$ was excluded from the
list of the best models (in accordance with COMPETE criteria, see details
in \cite{COMPETE}).

Analysis of these results shows that they are due to a poor description of
$\rho$ data at low energy. On the other hand, there are a few questions
concerning the explicit Regge-type models usually used for the analysis
and description of the data. How low in energy can the Regge
parame\-tri\-sations be extended, as they are written as functions of the
asymptotic variable $s$ rather than the ``Regge" variable $\cos\theta_{t}$
($=E/m$ in the laboratory system for identical colliding particles)? At
which energies can the ``asymptotic" normalization
\begin{equation}\label{eq:asympt. norm.}
\sigma_{tot}(s)=(1/s)\Im mA(s,0)
\end{equation}
instead of the standard one
\begin{equation}\label{eq:stand. norm.}
\sigma_{tot}(s)=(1/2mE)\Im mA(s,0)
\end{equation}
be used? And last, how much do the analytic expressions for $\rho$ based
on the derivative dispersion relations deviate from those calculated in
the integral form?

In this paper, we try to answer these questions considering three pomeron
models for $pp$ and $\bar pp$ interactions at $\sqrt{s}\geq 5$ GeV.

\section{Integral and derivative dispersion relations.}
Assuming, in accordance with many analyzes, that the odderon does not
contribute asymptotically at $t=0$, one can show that the integral
dispersion relations (IDR) for $pp$ and $\bar pp$ amplitudes can be
reduced to those with one subtraction constant \cite{rhohist}:
\begin{equation}\label{eq:dispers II}
\begin{array}{lll}
\rho_{\pm}\sigma_{\pm}&=&\frac{B}{2m_{p}p}\,\, +\\&+&\frac{E}{\pi p}\,{\rm
P}\!\int\limits_{m_{p}}^{\infty}\left
[\frac{\sigma_{\pm}}{E'(E'-E)}-\frac{\sigma_{\mp}}{E'(E'+E)}\right ]p'\,
dE'
\end{array}
\end{equation}
where $m_{p}$ is the proton mass, $E$ and $p$ are the energy and momentum
of the proton in the laboratory system, and $B$ is a subtraction constant,
usually determined from the fit to the data. The indices $+(-)$ stand
respectively for the $pp$ and $ (\bar pp)$ amplitudes. The standard
normalization (\ref{eq:stand. norm.}) is chosen in Eq.(\ref{eq:dispers
II}).

In the above expression, the contributions of the integral over the
unphysical cuts from the two-pion to the two-proton threshold are omitted
because they are $\lesssim 1\%$ (see, e.g. \cite{valeng}) in the region of
interest ($\sqrt{s}\ge 5$ GeV).

The derivative dispersion relations (DDR) were obtained \cite{DDR}
separately for crossing-even and crossing-odd amplitudes
\begin{equation}\label{eq:crossampl}
f_{\pm}(s,0)=A_{+}(s,0)\pm A_{-}(s,0).
\end{equation}
They are very useful in a practice due to their simple analytical form at
high energies, $E\gg m_{p}$:
\begin{equation}\label{eq:ddrev-as}
\Re ef_{+}(E,0) \approx E\tan \left [ \frac{\pi}{2}E\frac{d}{dE}\right ]
\Im mf_{+}(E,0)/E.
\end{equation}
However it is important to estimate the corrections to these asymptotic
relations (\ref{eq:ddrev-as}) if one is to use them at finite $s$.

It will be shown in the forthcoming paper that one can obtain for the even
part of amplitude
\begin{equation}\label{eq:ddreven}
\begin{array}{lll}
\Re ef_{+}(E,0)&=& B_{+}+E\tan \left [ \frac{\pi}{2}E\frac{d}{dE}\right
]\Im mf_{+}(E,0)/E\,
 -\\&-&\frac{2}{\pi
}\sum\limits_{p=0}^{\infty}\frac{C_{+}(p)}{2p+1}\left
(\frac{m_{p}}{E}\right )^{2p}
\end{array}
\end{equation}
with
$$
C_{+}(p)=\frac{e^{-\xi D_{\xi}}}{2p+1+D_{\xi}}[ \Im mf_{+}(E,0)-E\Im
mf'_{+}(E,0)].
$$
where $f'=df/dE$ and $D_{\xi}=d/d\xi, \xi=\ln(E/m_{p})$.

It can be proven, using the properties of $\exp(-\xi D_{\xi}),$ that
$C_{+}(p)$ does not depend on $E$. A similar expression is obtained for
the crossing-odd part of the amplitudes.

\section{Phenomenology.}
Our aim is to compare the fits of three pomeron models with $\rho(s)$
calculated by two methods: the integral dispersion relation, and the
asymptotic form of the DDR but with a subtraction constant.

We apply as a first step the IDR and DDR only to $pp$ and $\bar pp$ data,
fitting the high-energy models to the data at $\sqrt{s}\ge 5$ GeV.

\subsection{Low-energy data.}\label{sec:low}
Low-energy total cross sections for $pp$ (143 points) and $\bar pp$ (220
points) interactions at $\sqrt{s}<5$ GeV are explicitly parameterized and
the values of the free parameters are determined from a fit to the data at
$s<s_{min}$ (all data on $\sigma$ and $\rho$ are taken from \cite{data}).
Thus we perform an overall fit in three steps.
\begin{enumerate}
\item
 The chosen model for high-energy cross-sections is fitted to the data on
the cross sections only
at $s>s_{min}$.
 \item The obtained ``high-energy" parameters are fixed. The ``low-energy" parameters
are determined from the fit at $s<s_{min}$, but with $\sigma_{pp}^{\bar
pp}(s_{min})$  given by the first step.
 \item
The subtraction constant $B_{+}$ is determined from the fit at $s>s_{min}$
with all other parameters kept fixed.
\end{enumerate}

Then, without fitting, we calculate the ratios $\rho_{pp}$ and $\rho_{\bar
pp}$ at all energies above the physical threshold.

\subsection{High energy. Pomeron models.}\label{sec.high}

We consider three models leading to different asymptotic behaviors for the
total cross sections. We start from the explicit parameterization of the
total $pp$ and $\bar pp$ cross-sections, then, to find the ratios of the
real to imaginary parts, we apply the IDR making use of the second method
for a calculation of the low-energy part of the dispersion integral. Then
we compare results for ratios calculated through the DDR.

All the models include the contributions of pomeron, $f$ and $\omega$
reggeons (we consider these reggeons as effective ones because it is not
reasonable to add other secondary reggeons provided only the $pp$ and
$\bar pp$ data are fitted.)

\begin{equation}\label{eq:sigmod}
\sigma_{pp}^{\bar pp}={\cal P}(E)+R_{f}(E)\pm R_{\omega}(E),
\end{equation}
where $R_{f}=R_{+}, R_{\omega}=R_{-}$ and
\begin{equation}\label{eq:reggeon}
R_{\pm}(E)=g_{\pm}\left (E/\lambda m_{p}\right )^{\alpha_{\pm}(0)-1}.
\end{equation}
The parameter $\lambda$ can play a role only for the pomeron term in the
triple pole model (see below), in the simple pole and dipole models
$\lambda =1$.

When the imaginary part of amplitude is integrated in IDR we consider (for
comparison) two kinds of normalization: the standard one defined in Eq.
(\ref{eq:stand. norm.}) and the asymptotic one given by Eq.
(\ref{eq:asympt. norm.}). For the latter case, in the expressions for
cross-sections, the replacement $E/\lambda m_{p}\to s/s_{1}$ is made.

Besides, we compare our results with the models which are written as
functions of $-is$ and with the asymptotic normalization (\ref{eq:asympt.
norm.}).
\begin{equation}\label{-is}
\begin{array}{lll}
\frac{1}{s}A^{\bar pp}_{pp}(s,0)&=&i{\cal P}(-is)+iR_{f}(-is)\pm
R_{\omega}(-is)
\end{array}
\end{equation}
We denote such  fits as ``$-is$ fits". We present the results of the fits
using DDR (with a subtraction constant) and of standard fits with
amplitudes defined in accordance with ``$-is$" rule.

{\bf Simple pole pomeron model (SP).} In this model, the intercept of the
pomeron is larger than unity \cite{d-l}
\begin{equation}\label{simpleE}
{\cal P}(E)=g\left (E/(\lambda m_{p})\right )^{\alpha_{\cal P}(0)-1}.
\end{equation}

{\bf Dipole pomeron model (DP).} The pomeron in this model is a double
pole in the complex angular momentum plane with intercept $\alpha_{{\cal
P}}(0)=1$.
\begin{equation}\label{doubleE}
{\cal P}(E)=g_{1}+g_{2}\ln(E/\lambda m_{p}).
\end{equation}

{\bf Tripole pomeron model (TP)} The pomeron is the hardest complex
$j$-plane singularity allowed by unitarity: the triple pole at $t=0$ and
$j=1$.
\begin{equation}\label{tripoleE}
{\cal P}(E)=g_{1}+g_{2}\ln^{2}(E/\lambda m_{p}).
\end{equation}
\section{Results, discussion and conclusions.}

The quality of the fits is presented in the Table, where $\chi^{2}$ per
number of experimental points is shown. In the Figures we show the curves
only for the simple-pole pome\-ron model. For other models the curves are
practically the same in the region where data are available.

\vskip 0.7 cm
\begin{center}
\begin{tabular}{|c|c|c|c|c|} \hline
& \multicolumn{2}{c|}{Standard} & \multicolumn{2}{|c|}{Asymptotic}\\
& \multicolumn{2}{c|}{normalization} & \multicolumn{2}{|c|}{normalization}\\
\cline{2-5}
  & IDR & DDR & IDR & $"-is", B=0$ \\
\hline
 Simple Pole & 1.046 & 1.065 & 1.112 & 1.121 \\
\hline
 Double Pole & 1.053 & {1.045} & 1.132 & 1.079 \\
\hline
 Triple Pole & 1.044 & 1.046 & 1.111 & 1.115 \\
\hline
\end{tabular}
\end{center}
\noindent{\footnotesize Table. The values of the $\chi^{2}$ per point
($\chi^{2}/N_{p}$) obtained in the various pomeron models and through the
different methods for the calculation of the ratio~$\rho$. Both
$\sigma_{tot}$ and $\rho$ are included.}

\begin{figure}
\resizebox{0.4\textwidth}{!}{%
  \includegraphics{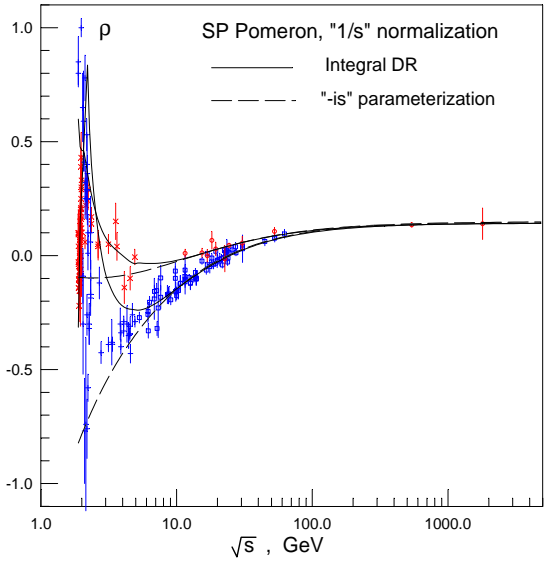}
}
\caption{The IDR fit with asymptotic normalization and the ``$-is$" fit.}
\label{fig:1}       
\end{figure}

\begin{figure}
\resizebox{0.4\textwidth}{!}{%
  \includegraphics{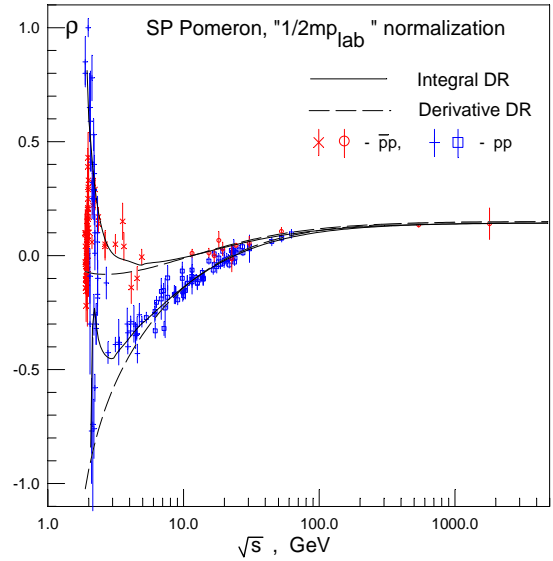}
}
\caption{The integral and derivative dispersion relation fit with the
standard normalization (see text for details).}
\label{fig:2}       
\end{figure}

As one can see from the Table, all models give similar descriptions of the
data on $\sigma_{tot}$ and $\rho$. Evidently, the fit with the integral
dispersion relations and with the standard normalization is preferable.
While the data on $\sigma$ are described with $\chi^{2}/N_{p} \approx
0.91$, the data on $\rho$ are described less well, with a
$\chi^{2}/N_{p}\approx 1.5$ in $pp$ case. We believe that this occurs
because of the bad quality of the $\rho$ data.

We would like to note that the values of $\rho$ calculated using DDR
deviate from those calculated with IDR even at $\sqrt{s} \lesssim 7 -8$
GeV (see Fig. 2). It means that in order to have more correct values of
the $\rho$ at such energies, one must use the IDR rather than explicit
analytical expressions from the DDR.

The neglect of the subtraction constant, together with the use of
asymptotic formulae in a non-asymptotic domain, may be the source of the
conclusion of \cite{COMPETE} excluding the simple-pole model from the list
of the best models. Inclusion of these (non-asymptotic) terms improves the
description of $\rho$ considerably, and may lead to different conclusions
regarding the simple-pole model.

However, in order to have these final conclusions, one will have to make a
complete (a la COMPETE) analysis of the all data, including cross sections
and $\rho$ for $\pi p$ and $Kp$ interactions\footnote{The first results
obtained in this direction (after the Conference) are given in
\cite{CLMS}.}.

\end{document}